\documentclass{article}
\usepackage{amssymb,amsmath,amsthm}
\newcommand{\ie}{\emph{ie}}
\newcommand{\etc}{\emph{etc}}
\newcommand{\cf}{\emph{cf}}
\newcommand{\Real}{\mathbb{R}}
\newcommand{\Nat}{\mathbb{N}}
\newcommand{\Hilbert}{\mathcal{H}}
\newcommand{\Dom}{\mathop{\mathrm{Dom}}\nolimits}
\newcommand{\si}{L^1}
\newcommand{\sii}{L^2}
\newcommand{\sobi}{\mathop{W_0^{1,2}}\nolimits}
\newcommand{\Sobi}{\mathop{W^{1,2}}\nolimits}
\newcommand{\Smooth}{C}
\newcommand{\demi}{\frac{1}{2}}

\newcommand{\dist}{\mathop{\mathrm{dist}}\nolimits}
\newcommand{\surface}{\mathcal{A}}
\newcommand{\curve}{\Sigma}
\newcommand{\strip}{\Omega}
\newcommand{\trans}{I}
\newcommand{\stripmap}{\mathcal{L}}
\newcommand{\tangent}{T}
\newcommand{\normal}{N}
\newcommand{\Hi}{\langle\mathsf{H}\mathrm{1}\rangle}
\newcommand{\Hii}{\langle\mathsf{H}\mathrm{2}\rangle}
\newcommand{\Hiii}{\langle\mathsf{H}\mathrm{3}\rangle}
\newcommand{\Hiv}{\langle\mathsf{H}\mathrm{4}\rangle}
\newcommand{\Hiiibis}{\langle\mathsf{H}\mathrm{3}'\rangle}
\newenvironment{Assumption}[1]
{\begin{description}\item[$#1$]\quad}
{\end{description}}
\newtheorem{Theorem}{Theorem}
\newtheorem{Corollary}{Corollary}
\theoremstyle{remark}
\newtheorem{Remark}{Remark}
\newenvironment{PROOF}{\begin{proof}[\textsc{Proof:}]}{\end{proof}}
\begin{document}
%
%
\title{
\textbf{Quantum Strips on Surfaces}
\footnote{PREPRINT}
}
\author{
David Krej\v{c}i\v{r}\'{\i}k
\footnote{
  On leave of absence from \newline
  \emph{Nuclear Physics Institute,
  Academy of Sciences,
  250\,68 \v{R}e\v{z} near Prague,
  Czech Republic}
\smallskip\newline
\textbf{Date:} April 25, 2002.
}
}
\date{{\small
\emph{
Laboratoire de Math\'ematiques,
Universit\'e de Reims, \\
Moulin de la Housse, BP~1039,
51\,687 Reims cedex~2, France \\
E-mail: david.krejcirik@univ-reims.fr
}}}
\maketitle
%
\begin{abstract}
\noindent
Motivated by the theory of quantum waveguides,
we investigate the spectrum of the Laplacian,
subject to Dirichlet boundary conditions,
in a curved strip of constant width
that is defined as a tubular neighbourhood
of an infinite curve in a two-dimensional Riemannian manifold.
Under the assumption that the strip is asymptotically
straight in a suitable sense, we localise the essential
spectrum and find sufficient conditions which
guarantee the existence of geometrically induced
bound states. In particular, the discrete spectrum exists
for non-negatively curved strips which are studied
in detail. The general results are used to recover
and revisit the known facts about quantum strips in the plane.
As an example of non-positively curved
quantum strips, we consider strips on ruled surfaces.
\end{abstract}
%
%
%
\section{Introduction}
The theory of quantum waveguides constitutes
a beautiful domain of mathematical physics
in which one meets an interesting interaction
of analysis and geometry.
Recall that the configuration space~$\strip$ of a waveguide
is usually modelled by tubular neighbourhoods of infinite
curves in~$\Real^d$, $d=2,3$ (quantum strips, tubes),
or surfaces in~$\Real^3$ (quantum layers),
while the dynamics is governed by the Laplace operator
with Dirichlet boundary conditions.
It is due to an admirable progress of mesoscopis physics
that such models do really represent actual nanostructures
which are produced in the laboratory nowadays.
We refer to~\cite{DE,LCM} for the physical background
and references.

A common, particularly interesting property of these systems
is that the curvature of the reference curve or surface may produce
bound states of the Laplacian below the essential spectrum.
This phenomenon was demonstrated first in a rigourous
way by P.~Exner and P.~\v{S}eba for curved strips
in the plane,~\cite{ES}. Numerous subsequent studies
improved their result and generalised it to space tubes.
For more information and other spectral and scattering
properties, see the review paper~\cite{DE} and references therein.
The evidently more complicated
case of quantum layers was investigated quite recently
in~\cite{DEK1,DEK2,EK3}.

Up to this time, the ambient manifold of the quantum waveguide
has been usually identified with a flat Euclidean
space~$\Real^d$, $d=2,3$. This restriction is obviously
due to the physical reasons, however, at least from the mathematical
point of view, one may be interested equally in the situations
when it is a general Riemannian manifold~$\surface$
of dimension~$d\geq 2$.
The principal interest of the present work
is to initiate this study by considering the simplest
non-trivial case, $d=2$, when the configuration space~$\strip$
is a tubular neighbourhood of constant radius~$a>0$
about an infinite curve~$\curve$ on a surface~$\surface$.

\medskip

Let us describe the contents of the paper.
The strip configuration space~$\strip$ itself is properly
defined in Section~\ref{Sec.Def}. Through all the paper,
we suppose that the strip is globally parameterised
by a system of geodesic coordinates based
on the reference curve~$\curve$.
In accordance with~\cite{Gray},
we call them \emph{Fermi coordinates},~\cite{Fermi},
although they had already been considered by C.~F.~Gauss.
A comprehensive discussion of such a coordinate system
has been given by F.~Fiala, \cite{Fiala}, in order
to prove some isoperimetric inequalities;
see also~\cite{Hartman64}.
A modern definition of Fermi coordinates of tubes about
a submanifold of a general Riemannian manifold
can be found in~\cite{Gray}.
We introduce them for our purposes in Section~\ref{Sec.Fermi}.

In Section~\ref{Sec.Hamiltonian},
the Hamiltonian~$H$ of our system is identified
with the Friedrichs extension of the Laplacian,
$-\Delta$ on~$\sii(\Omega)$,
which is expressed in Fermi coordinates and defined initially
on~$\Smooth_0^\infty(\Omega)$.
The construction is based on the quadratic-form approach
of~\cite[Chap.~6]{Davies}.
Two trivial classes of quantum strips are then mentioned
in Section~\ref{Sec.Flat}. If the curvature of the ambient
space is identically equal to zero on~$\strip$,
the strip is called \emph{flat}
and the spectrum of~$H$ coincides with the spectrum
of strips in the plane, \cite{DE}.
A generalisation of straight strips in the plane
is represented by \emph{geodesic} strips,
for which the reference curve is in addition a geodesic.
In that case, we find that the spectrum is
the interval~$[\kappa_1^2,\infty)$,
where~$\kappa_1:=\pi/(2a)$.

Section~\ref{Sec.Motivation} is devoted to a heuristic
analysis of of the Hamiltonian~$H$.
Using a unitary transformation, it can be identified
with a Schr\"odinger-like operator with a potential
expressed by means of the metric of~$\strip$.
The latter operator acquires a very instructive form
in the formal limit when the width of the strip tends
to zero. In particular, we reveal an effective potential
which is given by a combination of curvatures of~$\curve$ and~$\surface$.
The result is compared with the case of strips
in the plane and used as a motivation for the spectral analysis
of~$H$ in the subsequent sections.

In Section~\ref{Sec.EssSp}, we localise the essential spectrum
under the assumption that the strip is
\emph{asymptotically geodesic} in a suitable sense.
Using a Neumann bracketing argument together with
the minimax principle, we find in Theorem~\ref{EssSp}
that the threshold of~$\sigma_\mathrm{ess}(H)$
is then bounded from below by~$\kappa_1^2$.

Section~\ref{Sec.DiscSp} is devoted to the analysis
of the spectrum below the energy~$\kappa_1^2$.
Using a variational technique standard in the theory
of quantum waveguides,
we find two sufficient conditions which guarantee
that this part of spectrum is not empty,
\cf~Theorems~\ref{DiscSp1} and~\ref{DiscSp2}.
These conditions require that the strip is non-negatively
curved in an integral sense; see~(\ref{condition})
for the precise meaning of the statement.
Combining these results with Theorem~\ref{EssSp},
we arrive at Corollary~\ref{DiscSp} which contains
the main result of this paper concerning the existence
of a non-trivial discrete spectrum in quantum strips.

Since the condition~(\ref{condition})
is clearly satisfied for non-negatively curved strips,
this situation is investigated in detail
in Section~\ref{Sec.Positive}.
We simplify some assumptions, we have put on
the geometry of~$\strip$, and sum up the spectral
results in Theorem~\ref{Thm.Positive}.
Apart from a significant generalisation,
it recovers and revisits the known results
for the quantum strips in the plane.

To the best of our knowledge, it is for the first time when the
spectrum of a curved strip embedded in a non-trivial manifold has
been investigated.
An exception is the paper~\cite{CB1},
where I.~J.~Clark and A.~J.~Bracken
deal with a special class of quantum strips in~$\Real^3$,
which are made up from segments perpendicular
to an infinite space curve~$\Sigma$.
They introduce the Hamiltonian in a formal way,
derive the effective potential mentioned above
and make some conjectures on the influence
of the torsion of~$\curve$ on the spectrum, however,
do not perform any spectral analysis itself.
Actually, their paper is a preliminary to~\cite{CB2},
where bound states in space quantum waveguides
with torsion are investigated.
The strip of~\cite{CB1} is a part of a \emph{ruled}
surface~$\surface$ based on~$\curve$; we examine
this situation briefly in Section~\ref{Sec.Ruled}.

We conclude the paper by Section~\ref{Sec.Conclusions},
where some open problems and directions of a future research
are mentioned. A particularly interesting question
concerns possible applications to physics.

\section{Preliminaries}
%
\subsection{Definitions}\label{Sec.Def}
Let~$\surface$ be a non-compact two-dimensional Riemannian manifold
of class~$\Smooth^2$ and let~$K$ denote its Gauss curvature.
We require that~$K$ is a continuous function on~$\surface$,
which holds if~$\surface$ is of class~$\Smooth^3$
or if it is embedded in~$\Real^3$.
Even if it is not necessary for our construction,
we shall assume that~$\surface$ is geodesically complete.

Let~$\curve$ be a simple, infinite curve of class~$\Smooth^2$
embedded in~$\surface$ and let~$k$ denote its curvature.
(We do not require that~$\surface$ is embedded in~$\Real^3$,
however, if it is that case, $k$~means the geodesic curvature
of~$\curve$.)
We may assume that~$\curve$ is given by the image
of the mapping~$p:\Real\to\surface$ such that~$|p'|=1$.
It represents the $\Smooth^2$-parameterisation
of the curve by its arc length.
We note that~$k$ is a continuous function on~$\curve$.

Let~$a>0$ and~$\trans:=(-a,a)$.
The strip~$\strip$ of width~$2a$ is defined
as the $a$-tubular neighbourhood of~$\curve$ in~$\surface$:
\begin{equation}\label{DefStrip0}
  \strip:=\left\{ x\in\surface \:|\, \dist(x,\curve)<a \right\}.
\end{equation}
As usual, the distance~$\dist(x,\curve)$ means here
the length of the minimal geodesic joining~$x$ with~$\curve$.
We want to introduce the Laplacian in~$\strip$
and investigate its spectrum. Our strategy is to map
the curved strip~(\ref{DefStrip0})
onto the straight one, $\Omega_0:=\Real \times \trans$,
by the use of Fermi coordinates which are defined
in the following subsection.

\subsection{Fermi Coordinates}\label{Sec.Fermi}
We denote by~$\tangent_x\surface$ the tangent space to~$\surface$
at~$x\in\surface$ and recall that the exponential map,
$\exp_x:\tangent_x\surface \to \surface$,
is the identification~$t\mapsto\gamma_t(1)$,
where~$\gamma_t$ is the unique geodesic
(parameterised by arc length) in~$\surface$
with~$\gamma_t(0)=x$ and~$\gamma_t'(0)=t$.
We define
\begin{equation}\label{StripMap}
  \stripmap: \Real^2\to\surface:
  \left\{(s,u)\mapsto\exp_{p(s)}\left(u\,n(s)\right)
  \:|\, n\in\normal_p\curve\right\},
\end{equation}
where~$\normal_p\curve$ denote the orthogonal complement
of~$\tangent_p\curve$ in~$\tangent_p\surface$,
and always assume that
\begin{Assumption}{\Hi}
  $\stripmap: \strip_0 \to \strip$
  is a diffeomorphism for some~$a>0$.
\end{Assumption}
Then the inverse of~$\stripmap$ determines the system
of Fermi ``coordinates''~$(s,u)$ and one has
\begin{equation}\label{DefStrip}
  \strip=\stripmap(\strip_0) \,.
\end{equation}
\begin{Remark}
Hereafter we shall use the standard component notation of tensor
analysis with the range of indices being~$1,2$ and associate them
with Fermi coordinates via the identification,
$(1,2) \leftrightarrow (s,u)$. The partial derivatives
will be denoted by commas.
From now on the curvature~$K$ shall be considered
as a function of Fermi coordinates~$(s,u)$;
$k$~is a function of~$s$.
\end{Remark}

The metric tensor of~$\strip$ in Fermi coordinates
is given by $G_{ij}:=\langle\stripmap_{,i},\stripmap_{,j}\rangle$,
where~``$\langle\cdot,\cdot\rangle$'' denotes the inner product induced
by the Riemannian metric on~$\surface$.
Note that $s\mapsto\stripmap(s,u)$ traces the curves
parallel to~$\Sigma$ at a fixed distance~$|u|$
and that the curve $u\mapsto\stripmap(s,u)$ is a unit-speed
geodesic orthogonal to~$\Sigma$ for any fixed~$s$.
The generalised Gauss Lemma, \cite[Sec.~2.4]{Gray},
implies that these curves meet orthogonally
and one arrives at the diagonal form of the metric tensor
\begin{equation}\label{metric}
  (G_{ij})=
  \begin{pmatrix}
    f(s,u)^2 & 0 \\
    0        & 1
  \end{pmatrix}
  \,.
\end{equation}
According to~\cite{Hartman50,Hartman-Wintner},
the function~$f$ is continuous and has continuous
partial derivatives~$f_{,u}$, $f_{,uu}$
satisfying the Jacobi equation
\begin{equation}\label{Jacobi}
  f_{,uu}+K\,f=0
  \qquad\textrm{with}\qquad
  \begin{aligned}
    f(\cdot,0)&=1 \,, \\
    f_{,u}(\cdot,0)&=k \,.
  \end{aligned}
\end{equation}
The determinant of the metric tensor, $G:=\det(G_{ij})=f^2$,
defines through $d\strip:=G(s,u)^\demi ds du$
the surface element of the strip.

\begin{Remark}\label{Rem.Diff}
If~$\curve$ was a compact curve, then the condition~$\Hi$
could always be achieved for sufficiently small~$a$.
Recall also that the same holds true for infinite strips
in the plane if one assumes in addition that~$\strip$
does not overlap, \cite{DE}.
In our case, the situation is analogous.
The inverse function theorem implies that~$\stripmap:\strip_0\to\strip$
is a local diffeomorphism provided~$f$ is uniformly
strictly positive and bounded,
\cf~posterior assumption~$\Hii$.
This can be achieved for~$a$ small enough because~$f(\cdot,0)=1$.
The condition~$\Hi$ will be then fulfilled globally
if we do not allow in addition an overlapping of the strip.
\end{Remark}

We needed the ambient manifold~$\surface$ just in order to define
the strip by means of~(\ref{DefStrip0}) or~(\ref{DefStrip}).
Once the construction is over, we may forget about the rest
of~$\surface$ and consider its part~$\strip$ only.
It will be our configuration space.
Note that the closer~$\bar{\strip}$ is a manifold with boundary.

\subsection{Hamiltonian}\label{Sec.Hamiltonian}
After geometric preliminaries,
let us define the Hamiltonian of our system.
We consider a non-relativistic quantum particle
within the two-dimensional region~$\strip$
of impenetrable boundary.
As usual, we put~$\hbar^2/(2m)=1$, where~$\hbar$ denotes
Planck's constant and~$m$ the mass of the particle.
Then the Hamiltonian could be identified
with the Laplace operator, $-\Delta$ on~$\sii(\strip)$,
with an appropriate domain of functions
which vanish on~$\partial\strip$.
However, we proceed differently
and always understand this Laplacian
in the generalised (form) sense.

In detail, using Fermi coordinates, we shall identify
the Hilbert space~$\sii(\strip)$
with~$\Hilbert:=\sii(\strip_0,d\strip)$.
Let us consider the quadratic form on~$\Hilbert$ given by
\begin{equation}\label{form}
  Q(\psi,\phi):=
  \left( \psi_{,i},G^{ij}\phi_{,j} \right)_\Hilbert
  \,, \qquad
  \Dom Q:=\sobi(\strip_0,d\strip) \,,
\end{equation}
where~$(G^{ij})$ is the inverse of~$(G_{ij})$.
Assuming that the metric
is uniformly elliptic in the sense that the condition
\begin{Assumption}{\Hii}
  $\exists c_\pm>0 \quad \forall (s,u)\in\strip_0 \::
  \quad
  c_- \leq f(s,u) \leq c_+$
\end{Assumption}
is valid, it follows that the form~$Q$ is non-negative
and closed on its domain.
Consequently, there exists a non-negative self-adjoint
operator~$H$ associated to~$Q$
which satisfies~$\Dom H \subset \Dom Q$.
It will be our Hamiltonian.
We refer to~\cite[Chap.~6]{Davies} for more details
and proofs concerning the above construction.

\begin{Remark}
Although~$H$ is formally equal to the operator
$
  -G^{-\demi}\partial_i G^\demi G^{ij} \partial_j
$,
\ie\/ the Laplacian, $-\Delta$, expressed in Fermi coordinates,
we shall be particularly concerned
not to assume that the metric is differentiable.
If, however, the metric is sufficiently smooth
then the operator~$H$ is indeed given by this expression
with Dirichlet boundary conditions in the classical sense.
We stress that under our assumptions,
$f$ is only continuous w.r.t.~$s$.
\end{Remark}
%

\subsection{Flat and Geodesic Strips}\label{Sec.Flat}
Assume that the strip is \emph{flat} in the sense that the curvature~$K$
is equal to zero everywhere on~$\strip$, $K \equiv 0$.
Then the Jacobi equation~(\ref{Jacobi}) has the exact solution
\begin{equation}\label{FlatStrip}
  f(s,u)=1+u\,k(s) \,.
\end{equation}
This is a well-known result for the strips in the plane, however,
we note that the same holds as well for the strips on cylinders, on surfaces
of the shape of a corrugated iron, \etc.
Since the Hamiltonian is expressed via the metric
which depends on~$f$ only,
we may immediately adapt to the flat strips
all the results which has been previously derived
for the quantum strips in the plane,~\cite{DE}.
In particular, the discrete spectrum will always exist
as soon as the strip is non-trivially curved, $k \not\equiv 0$,
and asymptotically straight, $k \xrightarrow[]{\infty} 0$.

On the contrary, if (in addition to~$K \equiv 0$)
the reference curve is a geodesic, $k \equiv 0$,
then the function~$f$ equals~$1$ identically, and therefore
\begin{equation*}
  H=H_0:=-\Delta_D^{\strip_0}
  \qquad\textrm{on}\qquad
  \sii(\strip_0) \,.
\end{equation*}
Consequently, the discrete spectrum is empty and
\begin{equation}\label{GeodesicSpectrum}
  \sigma(H_0)=\sigma_\mathrm{ess}(H_0)
  =\left[\kappa_1^2,\infty\right) \,,
\end{equation}
where~$\kappa_1^2$ denotes the first eigenvalue
of the Dirichlet Laplacian on the transverse section,
$-\Delta_D^\trans$.
These systems generalises the straight strips in the plane
and will be called here \emph{geodesic}.
We will use them as a comparative class of quantum strips
whose spectrum is known explicitly.

The operator~$-\Delta_D^\trans$ occurs often in the present work.
We note that it is nothing else than the quantum Hamiltonian
of the one-dimensional infinite square well of width~$2a$.
In what follows we shall use its family
of eigenfunctions~$\{\chi_n\}_{n=1}^\infty$ which is given by
\begin{equation}\label{chi}
  \chi_n(u):=
\begin{cases}
  \sqrt{\frac{1}{a}} \, \cos\kappa_n u &\ \textrm{if \ $n$ is odd}, \\
  \sqrt{\frac{1}{a}} \, \sin\kappa_n u &\ \textrm{if \ $n$ is even},
\end{cases}
\end{equation}
where~$\kappa_n^2:=(\kappa_1 n)^2$ with~$\kappa_1:=\pi/(2a)$
are the corresponding eigenvalues.
The ground-state~$\chi_1$ will be very important
for us because it represents a generalised eigenvector
of the geodesic strip corresponding to the threshold
of the essential spectrum~(\ref{GeodesicSpectrum}).

\section{Motivation}\label{Sec.Motivation}
This part is devoted to heuristic considerations
in order to motivate the spectral analysis of the Hamiltonian
in the following sections. It is possible, but beyond
the scope of this paper, to examine the conditions
under which the thin-width limit process below is justified.
Since we use it just for motivation purposes,
we shall do the limit in a formal way only.
To this end (but only through this section!),
we shall assume that~$f$ is an analytic function.

Let us recall first the observation which initiated the attempts
to prove the existence of bound states in quantum strips in the plane,
\cf~\mbox{\cite{JK,KJ,daC1,Tol}}.
The Hamiltonian of such a strip is unitarily equivalent
to a Schr\"odinger-like operator with a potential expressed
by means of the curvature~$k$ of the reference curve and the transverse
coordinate~$u\in\trans$. Making formally
the thin-width limit, $a \to 0$, in the expression
for the the transformed Hamiltonian,
the potential becomes equal to~$-k^2/4$.
The latter always represents an \emph{attractive} interaction
as soon as the strip is non-trivially curved, $k \not\equiv 0$,
and asymptotically straight,~\mbox{$k \xrightarrow[]{\infty} 0$}.
Consequently, the limit operator possesses bound states
below its essential spectrum.
As we have mentioned in the introduction,
one proves that these bound states ``survive''
also in the actual quantum strips of non-zero widths.

In order to find the effective potential in our situation,
we introduce the unitary transformation
$U: \Hilbert \to \sii(\strip_0)$ given by
$\psi \mapsto G^\frac{1}{4}\psi$,
which leads to
\begin{equation}
  \tilde{H}:=U H U^{-1}=
  -G^{-\frac{1}{4}}\partial_i G^{\frac{1}{2}}
  G^{ij}\partial_j G^{-\frac{1}{4}} \,.
\end{equation}
Commuting~$G^{-\frac{1}{4}}$ with the gradient components,
we cast this operator into a form which has
a simpler kinetic part but contains a potential,
\begin{equation}\label{comHamiltonian}
  \tilde{H}=-\partial_i G^{ij}\partial_j+V
  \qquad\textrm{with}\qquad
  V:=(G^{ij}J_{,j})_{,i}+J_{,i}G^{ij}J_{,j} \,,
\end{equation}
where~$J:=\ln{G^\frac{1}{4}}$.
This expression is valid for any smooth metric~$G_{ij}$.
Employing the particular form~(\ref{metric}) of our metric tensor
together with the Jacobi equation~(\ref{Jacobi}), we get
\begin{equation}\label{potential}
  V = \frac{1}{f^2} \left[\frac{1}{2}\,\frac{f_{,ss}}{f}
  -\frac{5}{4}\left(\frac{f_{,s}}{f}\right)^2\right]
  -\frac{1}{2}\,K
  -\frac{1}{4} \left(\frac{f_{,u}}{f}\right)^2 .
\end{equation}

To make the limit when the width of the strip, $2a$, tends to zero,
we note first that the function~$f$
admits, as a solution of~(\ref{Jacobi}), the following
asymptotic expansion w.r.t.~$u$:
\begin{equation}
  f(s,u)=1+u\,k(s)-\mbox{$\demi$}\,u^2\,K(s,0)+r(s,u) \,,
\end{equation}
where the remainder~$r$ is~$\mathcal{O}(u^3)$ for any fixed~$s$.
Putting this expansion into~(\ref{potential})
and~(\ref{comHamiltonian}), and making the limit~$u \to 0$
in the expression for~$V$ and~$G^{ij}$, we see that,
up to higher order terms in~$u\in\trans$,
the operator~$\tilde{H}$ decouples formally
into the direct sum of the operators
\begin{equation}\label{thin-width}
  -\Delta^\Real+V_\mathrm{eff}
  \quad \textrm{on} \quad \sii(\Real)
  \qquad\textrm{and}\qquad
  -\Delta_D^\trans
  \quad \textrm{on} \quad \sii\left(\trans\right) \,,
\end{equation}
where
\begin{equation}\label{ThinPotential}
  V_\mathrm{eff}(s)
  :=-\mbox{$\frac{1}{4}$}\,k(s)^2-\mbox{$\demi$}\,K(s,0) \,.
\end{equation}
The first term in~$V_\mathrm{eff}$ is identical with
the effective potential for the thin strips in the plane,
while the second one reflects the fact that our strip
is in addition embedded in a curved manifold now.

Assume that the curvatures~$k$ and~$K$ vanish at the infinity
of the strip. In distinction to the planar case,
the potential~(\ref{ThinPotential}) may not represent
an attractive interaction for any non-trivially curved strip.
(For, it suffices to consider the strip constructed
over a geodesic, $k \equiv 0$, on a surface of negative
curvature.) Nevertheless, if the curvature~$K$ is, say, non-negative
(and $k \not\equiv 0$ provided~$K \equiv 0$),
then the potential~$V_\mathrm{eff}$ always represents
an attractive interaction. Consequently, the direct sum
of the limit operators of~(\ref{thin-width}) possesses bound states
below its essential spectrum. The aim of this paper is to state
an analogous sufficient condition which guarantees
the existence of a non-trivial discrete spectrum
for the actual Hamiltonian~$H$ of the strips of non-zero widths.

To conclude this section, we stress again that the thin-width-limit
procedure we have used to derive the operators~(\ref{thin-width})
and the effective potential~(\ref{ThinPotential}) of thin strips
is formal only. (One reason is that the transverse
operator~$-\Delta_D^\trans$ gives rise to infinite normal
oscillations as~$a \to 0$.) Nevertheless, we note that
a similar thin-neighbourhood limit was performed rigourously
by R.~Froese and I.~Herbst,~\cite{FrHe}, in efforts to treat
the time evolution around a compact $n$-dimensional submanifold
of~$\Real^{n+m}$, $m \geq 1$. There the confinement was realised
by a harmonic potential transverse to the manifold and the limit
was carried out by means of a dilation procedure followed
by averaging in the normal direction.
The situation when~$\Real^{n+m}$ is replaced by a Riemannian manifold
of the same dimension was treated formally in~\cite{Mitchell};
there one can also recover the effective potential~(\ref{ThinPotential}).

\section{Essential Spectrum}\label{Sec.EssSp}
In Section~\ref{Sec.Flat}, we have seen that the essential spectrum
of a geodesic strip (\mbox{$k,K \equiv 0$}) starts by the first
eigenvalue~$\kappa_1^2$ of the transverse operator~$-\Delta_D^\trans$.
Since the metric tensor is the identity matrix ($f \equiv 1$)
in this case and the essential spectrum is determined
by the behaviour of the metric at infinity only,
we expect that the same will hold true if a curved quantum strip
behaves like a geodesic strip \emph{asymptotically}
in the sense
\begin{Assumption}{\Hiii}
$
  f\xrightarrow[]{\ \infty\ }1 \,.
$
\end{Assumption}
By the symbol~``$\xrightarrow[]{\infty}$'' we mean precisely
the limit as~$s_0$ tends to~$+\infty$ from the supremum
over~$(s,u) \in \strip_0\setminus\strip_{0,\mathrm{int}}$,
where~$\strip_{0,\mathrm{int}}:=(-s_0,s_0)\times\trans$.
\begin{Remark}
Note that the assumption~$\Hiii$ together with~$\Hi$ implies
the condition~$\Hii$ for any half-width less than~$a$.
In detail, since~$f$ is continuous it is bounded locally,
and cannot be equal to~$0$ on~$\strip_0$ because of~$\Hi$.
The asymptotic assumption~$\Hiii$ then controls the uniform
behaviour of~$f$ at infinity.
\end{Remark}
\begin{Theorem}\label{EssSp}
Assume~$\Hi$, $\Hii$, and suppose that the strip is asymptotically
geodesic,~$\Hiii$. Then
$$
  \inf\sigma_\mathrm{ess}(H) \geq \kappa_1^2 \,.
$$
\end{Theorem}
\begin{PROOF}
The idea is inspired with the proof of Theorem~4.1 in~\cite{DEK2}.
Let~\mbox{$s_0>0$}, be~$\strip_{0,\mathrm{int}}$ as above and define
$\strip_{0,\mathrm{ext}}:=\strip_0\setminus\strip_{0,\mathrm{int}}$.
The images of~$\strip_{0,\mathrm{int}}$
and~$\strip_{0,\mathrm{ext}}$ by the mapping~$\stripmap$
divide the strip~$\strip$ into an interior and exterior
part, respectively.
Imposing the Neumann boundary condition
at the common boundary of the two parts, $s=s_0$, we arrive
at the decoupled Hamiltonian
$H^N=H_\mathrm{int}^N\oplus H_\mathrm{ext}^N$.
More precisely, it is obtained as the operator associated
with the quadratic form~$Q^N$ acting as~(\ref{form}), however,
with the domain
$\Dom Q^N:=\Dom Q_\mathrm{int}^N\oplus\Dom Q_\mathrm{ext}^N$,
where
$$
  \Dom Q_\omega^N
  :=\left\{\psi\in\Sobi(\strip_{0,\omega},d\strip)\,|\
  \psi(\cdot,\pm a)=0\right\} \,,
  \qquad
  \omega\in\{\mathrm{int},\mathrm{ext}\} \,.
$$
The corresponding quadratic forms~$Q_\mathrm{\omega}^N$
act like~$Q$, however, on appropriately restricted
Hilbert spaces~$\Hilbert_\mathrm{\omega}:=\sii(\strip_{0,\omega},d\strip)$.
Since $H\geq H^N$ and the spectrum of~$H_\mathrm{int}^N$
is purely discrete, \cite[Chap.~7]{Davies},
the minimax principle, \cite[Sec.~XIII.~1]{RS4},
gives the estimate
$$
  \inf\sigma_\mathrm{ess}(H)\geq\inf\sigma_\mathrm{ess}(H_\mathrm{ext}^N)
  \geq\inf\sigma(H_\mathrm{ext}^N) \,.
$$
Hence it is sufficient to find
a lower bound on~$H_\mathrm{ext}^N$.
However, by virtue of~(\ref{form}) and~(\ref{metric}),
we have for all~$\psi\in\Dom Q_\mathrm{ext}^N$:
\begin{eqnarray*}
\lefteqn{Q_\mathrm{ext}^N[\psi]
  \ \geq \
  \|\psi_{,u}\|_{\Hilbert_\mathrm{ext}}^2
  \ \geq \
  \big(\inf_{\strip_{0,\mathrm{ext}}} \!\! G^\demi \big) \
  \|\psi_{,u}\|_{\sii(\strip_{0,\mathrm{ext}})}^2} \\
&&  \ \geq \
  \big(\inf_{\strip_{0,\mathrm{ext}}} \!\! G^\demi \big) \
  \,\kappa_1^2\ \|\psi\|_{\sii(\strip_{0,\mathrm{ext}})}^2
  \ \geq \
  \big(\inf_{\strip_{0,\mathrm{ext}}} \!\! G^\demi \big)
  \big(\sup_{\strip_{0,\mathrm{ext}}} \! G^\demi \big)^{-1} \,
  \kappa_1^2 \ \|\psi\|_{\Hilbert_\mathrm{ext}}^2 \,.
\end{eqnarray*}
In the third inequality, we have used
the bound~$-\Delta_D^\trans \geq \kappa_1^2$.
The obtained estimate on~$Q_\mathrm{ext}^N$
is valid for any metric of the block form~(\ref{metric})
even if the function~$f$ is replaced by a matrix.
However, here we have~$G^\demi=f$ and the infimum and supremum
tend to~$1$ as~$s_0\to\infty$ by the assumption~$\Hiii$.
The claim then easily follows by the fact
that~$s_0$ can be chosen arbitrarily large.
\end{PROOF}
\begin{Remark}
This threshold estimate is sufficient for the subsequent
investigation of the discrete spectrum which is our goal in this paper.
In order to prove the opposite estimate, one may employ
a Dirichlet bracketing argument instead of the Neumann one we have used.
Next, to show that all energies above~$\kappa_1^2$ belong
to the spectrum, one has to construct an appropriate Weyl sequence.
This can be done under an assumption stronger than~$\Hiii$
which involves derivatives of~$f$ as well.
\end{Remark}
%

\section{Discrete Spectrum}\label{Sec.DiscSp}
The aim of this section is to prove two conditions
sufficient for the Hamiltonian to have a non-empty spectrum
below~$\kappa_1^2$. Since we have shown that the essential spectrum
does not start below this value for the asymptotically geodesic strips,
the conditions yields immediately the existence of curvature-induced
bound states. The proofs here are based on the variational idea
of finding a trial function~$\psi$ from the form domain of~$H$ such that
\begin{equation}
  \Tilde{Q}[\psi]:=Q[\psi]-\kappa_1^2\,\|\psi\|_\Hilbert^2<0 \,.
\end{equation}

The idea which goes back to J.~Goldstone and R.~L.~Jaffe, \cite{GJ},
is to construct a trial function by deforming~$\chi_1$ of~(\ref{chi}),
which represents a generalised eigenfunction of energy~$\kappa_1^2$
for the geodesic strip.
In particular, if the strip is geodesic, then~$\tilde{Q}[\chi_1]=0$.
The latter has to be understood in a generalised sense
because~$\chi_1$ is not integrable w.r.t.~$s$ and as such
it does not belong to~$\Dom Q$. Let us use this function
in the curved case. We start with a formal calculation:
\begin{eqnarray}\label{FormalResult}
  \tilde{Q}[\chi_1]
&=& \left(\chi_{1,s},f^{-1} \chi_{1,s}\right)
  + \left(\chi_{1,u},f \chi_{1,u}\right)
  -\kappa_1^2 \left(\chi_1,f\chi_1\right) \nonumber \\
&=& -\left(\chi_1,f_{,u}\,\chi_{1,u}\right)
  =\mbox{$\frac{1}{2}$}\,\left(\chi_1,f_{,uu}\,\chi_1\right)
  =-\mbox{$\frac{1}{2}$}\,\left(\chi_1,K f \chi_1\right) \,,
\end{eqnarray}
where the inner product is in the Hilbert space~$\sii(\strip_0)$.
The first equality is the definition of~$Q$ and~$\|\cdot\|_\Hilbert$,
in the second one we have used the fact that~$\chi_1$ does not depend
on~$s$ and integrated by parts w.r.t.~$u$,
in the third one we have integrated by parts once more,
and the last equality follows by~(\ref{Jacobi}).
The resulting integral will be well defined if we assume
\begin{Assumption}{\Hiv}
$
  K \in \si(\strip_0,d\strip) \,.
$
\end{Assumption}
Hence we obtain immediately
\begin{Theorem}\label{DiscSp1}
Assume~$\Hi$, $\Hii$, $\Hiv$,
and suppose that
\begin{equation}\label{K>0}
  \left(\chi_1,K \chi_1\right)_\Hilbert > 0 \,.
\end{equation}
Then
$$
  \inf\sigma(H) < \kappa_1^2 \,.
$$
\end{Theorem}
\begin{PROOF}
It remains to regularise~$\chi_1$ in such a way that
the formal result~(\ref{FormalResult}) would be justified in a limit.
For any~$n\in\Nat\setminus\{0\}$, we define
$\psi_n:=\varphi_n\chi_1$, where, for example,
$$
  \varphi_n(s):=
  \begin{cases}
    1 & \ \textrm{if} \quad |s|\in [0,n), \\
    (2n-|s|)/n & \ \textrm{if} \quad |s|\in[n,2n), \\
    0 & \ \textrm{if} \quad |s|\in [2n,\infty).
  \end{cases}
$$
Although~$\psi_n$ is not smooth, it is continuous and as such
it as an admissible trial function from~$\Dom Q$.
Since the variables~$(s,u)$ are separated in~$\psi_n$,
we arrive easily at
\begin{equation}\label{Result}
  \tilde{Q}[\psi_n]
  = \left(\psi_{n,s},f^{-1} \psi_{n,s}\right)
  -\mbox{$\frac{1}{2}$}\,\left(\psi_n,K f \psi_n\right) \,,
\end{equation}
where the first term vanishes as~$n\to\infty$ because
$$
  0<\left(\psi_{n,s},f^{-1} \psi_{n,s}\right)
  \leq c_-^{-1} \|\varphi_n'\|_{\sii(\Real)}^2
  =\frac{2c_-^{-1}}{n} \,.
$$ We have employed here~$\Hii$ and the normalisation of~$\chi_1$.
Since~$\varphi_n \to 1$ point-wise as~$n\to\infty$ and $K$~is
supposed to be integrable, the second term in~(\ref{Result})
converges to the negative integral of~(\ref{FormalResult})
by the dominated convergence theorem.
Consequently, there exists a fixed~$n_0$
such that~$\tilde{Q}[\psi_{n_0}]$ is negative
and the proof is finished.
\end{PROOF}
It may not be easy to verify the sufficient condition~(\ref{K>0}) for
a given ambient surface~$\surface$ and reference curve~$\Sigma$.
Nevertheless, it is clear that it holds true for any strip of
positive curvature. On the other hand, the condition is not satisfied
for the strips in the plane where, however, it is well known
that any non-trivial curvature of~$\Sigma$ pushes the spectrum of~$H$
below the energy~$\kappa_1^2$. The following result shows that
the same holds true for a more general class of quantum strips,
including the flat case too.

\begin{Theorem}\label{DiscSp2}
Assume~$\Hi$, $\Hii$, $\Hiv$,
and suppose that
\begin{equation}\label{K=0}
  \left(\chi_1,K \chi_1\right)_\Hilbert = 0 \,.
\end{equation}
If~$K \equiv 0$, we require in addition that~$k \not\equiv 0$.
Then
$$
  \inf\sigma(H) < \kappa_1^2 \,.
$$
\end{Theorem}
\begin{PROOF}
Let us start with formal considerations.
By virtue of~(\ref{FormalResult}),
the condition~(\ref{K=0}) implies that $\tilde{Q}[\chi_1]=0$.
It is the result which one obtains for the strips in the plane.
There the usual strategy is to deform slightly the function~$\chi_1$
on a curved part of the strip in order to obtain
a negative value of the functional~$\tilde{Q}$.
In particular, let~$\varepsilon\in\Real$
and there exist a function~$\phi$ of a compact support
in~$\strip_0$ such that it belongs to~$\Dom Q$
and $\tilde{Q}(\phi,\chi_1)$ is not equal to zero.
Writing
$$
  \tilde{Q}[\chi_1+\varepsilon\phi]
  =\tilde{Q}[\chi_1]+2\varepsilon\,\tilde{Q}(\phi,\chi_1)
  +\varepsilon^2\tilde{Q}[\phi] \,,
$$
and since the first term at the r.h.s. equals zero,
we can choose~$\varepsilon$ sufficiently small
and of a suitable sign so that the sum of the last two terms
is negative. The result is then justified by using
the mollifier~$\varphi_n$ from the proof of the previous theorem
in order to regularise~$\chi_1$.
Since the function~$\varphi_n$ equals one on an interval
growing as~$n\to\infty$ and~$\phi$ is of a compact support,
we can take~$n$ sufficiently large so that
$\tilde{Q}(\phi,\varphi_n\chi_1)$ does not depend on~$n$.
Hence it suffices to find an appropriate function~$\phi$
which verifies the above properties.

If~$K \not\equiv 0$, we take~$\phi(s,u):=j(s,u)^2\chi_1'(u)$,
where~$j$ is a non-zero infinitely smooth function
with a compact support on a region in~$\strip_0$
where~$f_{,u}$ does not change sign and it is not identically zero.
Such a region surely exists because~$f_{,u}$ is a continuous
function satisfying~(\ref{Jacobi}).
Then
$$
  \tilde{Q}(\phi,\chi_1)
  =-\left(j\chi_1',f_{,u} j\chi_1'\right)
  \not= 0 \,.
$$

If~$K \equiv 0$, we take~$\phi(s,u):=j(s)^2 u \chi_1(u)$,
where~$j$ is a non-zero infinitely smooth function
with a compact support on an interval in~$\Real$
where~$k$ does not change sign and it is not identically zero.
Then
$$
  \tilde{Q}(\phi,\chi_1)
  =\mbox{$\frac{1}{2}$}\,\left(j\chi_1,f_{,u} \, j\chi_1\right)
  =\left(j,kj\right)_{\sii(\Real)}
  \not=0 \,,
$$
where we have used the explicit form~(\ref{FlatStrip}) of~$f$
for the flat strips.
\end{PROOF}
\begin{Remark}
If~$K \equiv 0$, we have already mentioned that
the idea of the proof belongs to~\cite{GJ}.
Nevertheless, the deformation is not given explicitly there.
An explicit deformation function is used in~\cite{DE},
however, we have not used it here because it would require
an extra condition on the regularity of~$f$.
Our trial function is inspired with~\cite{RB};
see also~\cite[Thm.~5.1]{DEK2}.
\end{Remark}

An immediate consequence of Theorems~\ref{EssSp},
\ref{DiscSp1} and~\ref{DiscSp2} is the following
\begin{Corollary}\label{DiscSp}
Assume~$\Hi$, $\Hii$, $\Hiii$, $\Hiv$, and suppose that
\begin{equation}\label{condition}
  \left(\chi_1,K \chi_1\right)_\Hilbert \geq 0 \,.
\end{equation}
If~$K \equiv 0$, we require in addition that~$k \not\equiv 0$.
Then
$$
  \sigma_\mathrm{disc}(H) \not= \emptyset \,,
$$
\ie, there exists at least one isolated eigenvalue
of finite multiplicity situated below~$\kappa_1^2$.
\end{Corollary}
%

\section{Non-Negative Curvature}\label{Sec.Positive}
Since the condition~(\ref{condition})
is clearly satisfied for non-negatively curved strips,
we shall suppose that~$K \geq 0$ through all this section
and investigate this situation in detail.
Since the integral $\left(\chi_1,K \chi_1\right)_\Hilbert$
is always well defined, we may not assume the assumption~$\Hiv$.
This includes to use the monotone convergence theorem
instead of the dominated one we have
used in the proofs of Theorems~\ref{DiscSp1} and~\ref{DiscSp2}.

An integration of the Jacobi equation~(\ref{Jacobi})
yields the following identity
\begin{equation}\label{JacobiIntegrated}
  \forall (s,u)\in\strip_0:
  \qquad
  f_{,u}(s,u) = k(s)-\int_0^u K(s,\xi)\,f(s,\xi)\,d\xi \,.
\end{equation}
Since~$K$ is non-negative, we have immediately
\begin{equation}\label{f<}
  f(s,u) \leq 1+u\,k(s) \,.
\end{equation}
Let~$a\|k\|_\infty<1$.
Putting the inequality~(\ref{f<}) into~(\ref{JacobiIntegrated}),
we get an opposite bound
\begin{equation}\label{f>}
  f(s,u) \geq 1+u\,k(s)
  -\mbox{$\demi$}\,u^2
  \left(1+\mbox{$\frac{1}{3}$}\,u\,k(s)\right)
  \,\sup_{\xi\in\trans} K(s,\xi)
  \,.
\end{equation}
It follows from~(\ref{f<}) and~(\ref{f>}) that the condition~$\Hii$
can always be achieved for bounded curvatures and~$a$ small enough.
More specifically, a condition on the half-width is expressed
by means of the following inequality
\begin{equation}\label{maxa}
  \frac{1}{6}\,a^2\,\|K\|_\infty
  +\frac{2}{3-a\|k\|_\infty}
  \,<\,1 \,.
\end{equation}
(The supremum norm of~$K$ is taken over the strip only.)
We note that the condition $a\|k\|_\infty<1$ is a usual
assumption in the theory of quantum strips in the plane,
while the presence of~$K$ in~(\ref{maxa}) is due to the curved
ambient space~$\surface$.

Furthermore, it is clear from~(\ref{f<}) and~(\ref{f>})
that the asymptotic condition~$\Hiii$ is satisfied
if we assume
\begin{Assumption}{\Hiiibis}
$
  k \xrightarrow[]{\ \infty\ } 0
  \qquad\textrm{and}\qquad
  K \xrightarrow[]{\ \infty\ } 0 \,.
$
\end{Assumption}
We refer to the beginning of Section~\ref{Sec.EssSp}
for the exact definition of~``$\xrightarrow[]{\infty}$''.
The first limit is the usual assumption on the asymptotic
straightness of the strips in the plane, while the second
requires that the surface~$\strip$ is asymptotically flat.
The latter restricts the asymptotic behaviour of the ambient
space~$\surface$.

Finally, we remind that also the basic assumption~$\Hi$ can
always be a\-chieved for sufficiently small~$a$
if one assumes in addition that the strip does not overlap,
\cf~Remark~\ref{Rem.Diff}.
Summing up the above considerations together with the results
of the precedent sections, we conclude by
\begin{Theorem}\label{Thm.Positive}
Let~$\strip$ be a strip of non-negative curvature, $K \geq 0$,
which does not overlap and satisfies the condition~\emph{(\ref{maxa})}
together with~$a\|k\|_\infty<1$.
If it is not a geodesic strip, $k \not\equiv 0$ or $K \not\equiv 0$,
then
$
  \inf\sigma(H) < \kappa_1^2.
$
If it is in addition an asymptotically geodesic strip, $\Hiiibis$,
then the essential spectrum starts above~$\kappa_1^2$
and~$H$ has at least one isolated eigenvalue of finite multiplicity.
\end{Theorem}

This theorem generalises the known results for strips in the
plane, \cite{DE}, which are a particular case of the flat strips,
$K\equiv 0$. Moreover, the condition which enables us to localise the
essential spectrum is weaker in the sense that it does not contain
derivatives of the curvature~$k$ of the reference curve.
However, the most important generalisation concerns the quantum
strips on non-trivially curved manifolds
with a positive curvature. An instructive example in~$\Real^3$
is given by the infinite strips on the paraboloid of revolution.

\section{Ruled Strips}\label{Sec.Ruled}
In Section~\ref{Sec.Flat}, we have found an explicit form of the
metric~(\ref{metric}) in Fermi coordinates for the flat strips which
represent a trivial situation ($K \equiv 0$).
In general, however, it is not at all an easy problem to find~$f$
because it requires to determine the geodesics
orthogonal to the reference curve~$\curve$ and integrate the Jacobi
equation~(\ref{Jacobi}) over these geodesics.
Nevertheless, there is a non-trivial class of strips in~$\Real^3$
where the metric is easy to calculate.
For, consider the strip~$\strip$ constructed by segments
orthogonal to a space curve~$\curve$.
Such a strip is a part of a \emph{ruled} surface~$\surface$
based on~$\curve$, \cite[Def.~3.7.4]{Kli}.
As we have mentioned in the introduction,
the Hamiltonian~$H$ of a quantum particle in the ruled strips
had already been investigated in~\cite{CB1}.
The aim of the present paper is just to derive another
expression for~$f$, which suits better to our approach,
and discuss some properties of~$H$.
A more detailed spectral analysis of the ruled strips
will be discussed elsewhere.

Let~$\curve$ be a simple, infinite curve of class~$\Smooth^3$
in~$\Real^3$ and $p:\Real\to\Real^3$ be its parameterisation
by the arc length~$s$.
We assume that the set~$\{\dot{p},n,b\}$,
where~$n$ and~$b$ are the unit normal and binormal vectors,
respectively, is well defined and forms a right-handed Frenet
triad frame. We use the symbols~$\kappa$ and~$\tau$ for the curvature
and torsion of~$\Sigma$, respectively.
One general class of ruled surfaces~$\surface$ is defined via
$\stripmap: \Real^2\to\Real^3$,
\begin{equation}\label{RuledMap}
  \stripmap(s,u):=
  p(s)+u\left[n(s)\cos\theta(s)-b(s)\sin\theta(s)\right]
  \,,
\end{equation}
where~$\theta:\Real\to\Real$ is a function of class~$\Smooth^1$.
The ruled strip~$\strip$ is then given by~(\ref{DefStrip})
so that~$\Hi$ and~$\Hii$ hold true.
The mapping~(\ref{RuledMap}) does really represent the Fermi-coordinate
chart~(\ref{StripMap}) with the metric of the form~(\ref{metric}).
Employing the Frenet-Serret formulae, an explicit calculation yields
\begin{eqnarray}
  f(\cdot,u)^2
&=& \left(1-u\,\kappa\cos\theta\right)^2+u^2(\tau-\theta')^2
  \label{Ruled1} \\
  K(\cdot,u)
&=&
  -\frac{(\tau-\theta')^2}{f(\cdot,u)^2} \ ,
  \qquad\quad
  k\ =\ -\kappa\cos\theta \,.
  \label{Ruled2}
\end{eqnarray}
It is clear that any ruled strip has always a non-positive curvature.
Consequently, the sufficient condition~(\ref{condition})
is achieved only in the limit case,
$K \equiv 0$, which corresponds to~$\theta'=\tau$.
In that case, $\Omega$ is a flat strip
which may not be necessary a part of plane, however.

Combining~(\ref{Ruled1}) with~(\ref{Ruled2}), we get
\begin{equation}\label{Ruled}
  f(s,u)
  =\frac{1+u\,k(s)}{\sqrt{1+u^2K(s,u)}} \ ,
\end{equation}
which is an expression of a more transparent structure
from the intrinsic point of view of this paper.
At the same time, it is clear from~(\ref{Ruled})
that the condition~$\Hii$ holds true provided
\begin{equation}
  a \|k\|_\infty < 1
  \qquad\textrm{and}\qquad
  a^2 \|K\|_\infty <1 \,,
\end{equation}
and the assumption~$\Hi$ then follows by the additional
requirement that~$\strip$ does not overlap.
Next, the ruled strip is asymptotically geodesic under
the assumption~$\Hiiibis$, which implies that
$
  \inf\sigma_\mathrm{ess}(H) \geq \kappa_1^2
$
by Theorem~\ref{EssSp}.
However, an open question is whether there exist
bound states below the threshold of the essential spectrum
provided~$K \not\equiv 0$.

\section{Concluding Remarks}\label{Sec.Conclusions}
The main interest of this paper was to investigate spectral
properties of the Laplacian~$-\Delta$, subject to Dirichlet boundary
conditions, in the strip region~$\strip$ defined
as the tubular neighbourhood of an infinite curve~$\curve$
in a two-dimensional Riemannian manifold~$\surface$.
The strategy was to express the operator~$-\Delta$
under suitable assumptions, $\Hi$, $\Hii$,
in geodesic coordinates based on~$\curve$.
We were mainly interested in the existence of the discrete
spectrum. In particular, using some variational techniques,
we proved that there are bound states below the essential
spectrum provided the strip
is not geodesic, $K \not\equiv 0$ or $k \not\equiv 0$,
but asymptotically geodesic, $\Hiii$,
and positively curved ``in the mean''
in the sense of~(\ref{condition}).
The latter sufficient conditions hold particularly true
for the strips of a non-negative curvature which were investigated
in detail. The obtained results represent a generalisation
of quantum strips in the plane, \cite{DE}.

An interesting problem is to decide whether the discrete spectrum
exists for some negatively curved quantum strips as well.
The simplest model is probably given by the ruled strips
introduced in the previous section.
It is also desirable to investigate the spectrum of quantum strips
on surfaces more precisely using some perturbation and numerical methods.
Another direction of a future research consists in quantum
strips which are not asymptotically geodesic; this may include
periodically or randomly curved strips too.
Following~\cite{DKriz2}, we also expect that interesting new features
may be brought by a switch of the boundary condition.
Apart from the spectral analysis,
the scattering problem represents another challenge
facing the theory of quantum strips.

The present paper has been motivated by the theory of quantum
waveguides. If one deals with a curved quantum waveguide in the plane,
a reasonable model is given by the two-dimensional Laplacian in an
infinite strip in~$\Real^2$, \cite{DE}. However, we stress here that
the two-dimensional Laplacian in the strip on a curved surface
does not represent the actual Hamiltonian of a space quantum waveguide.
For, a quantum particle in a strip-like waveguide
is forced to move close to~$\strip$ by means of a constraining
potential (representing a high chemical potential between
different semiconductor materials) but, due to tunneling effect,
it can be found, even if not too far, outside the strip in
the space~$\Real^3$ too.
Even if this effect is not important for the waveguide in the plane
because the motion of the particle in the direction transverse
to the plane can be separated,
it is not negligible for waveguides on a curved surface.
In this paper, we dealt with a more general
situation when the ambient space~$\surface$ of the waveguide
may not be embedded in~$\Real^3$.
Our results are interesting from the mathematical
point of view, however, it is worth to know whether they
could be interpreted physically as well.

\section*{Acknowledgments}
The author would like to thank Professors
Pierre Duclos, Pavel Exner, and Jean Nourrigat
for useful discussions.
The work has been partially supported by
GA AS\,CR grant IAA 1048101.

%

\begin{thebibliography}{10}

\bibitem{CB2}
I.~J. Clark and A.~J. Bracken, \emph{Bound states in tubular quantum waveguides
  with torsion}, J.~Phys.~A \textbf{29} (1996), 4527--4535.

\bibitem{CB1}
\bysame, \emph{Effective potentials of quantum strip waveguides and their
  dependence upon torsion}, J.~Phys.~A \textbf{29} (1996), 339--348.

\bibitem{daC1}
R.~C.~T. da~Costa, \emph{Quantum mechanics of a constrained particle}, Phys.
  Rev. A~ \textbf{23} (1981), 1982--1987.

\bibitem{Davies}
E.~B. Davies, \emph{Spectral theory and differential operators}, Camb. Univ
  Press, Cambridge, 1995.

\bibitem{DKriz2}
J.~Dittrich and J.~K\v{r}{\'\i}\v{z}, \emph{Curved planar quantum wires with
  {D}irichlet and {N}eumann boundary conditions}, to appear in J.~Phys.~A.
  E-preprint: \textsf{mp\_arc~02-100} or \textsf{math-ph/0203007}.

\bibitem{DE}
P.~Duclos and P.~Exner, \emph{{C}urvature-induced bound states in quantum
  waveguides in two and three dimensions}, Rev. Math. Phys. \textbf{7} (1995),
  73--102.

\bibitem{DEK1}
P.~Duclos, P.~Exner, and D.~Krej\v{c}i\v{r}\'{\i}k, \emph{Locally curved
  quantum layers}, Ukrainian J.~Phys. \textbf{45} (2000), 595--601.

\bibitem{DEK2}
\bysame, \emph{Bound states in curved quantum layers}, Commun. Math. Phys.
  \textbf{223} (2001), 13--28.

\bibitem{EK3}
P.~Exner and D.~Krej\v{c}i\v{r}\'{\i}k, \emph{Bound states in mildly curved
  layers}, J.~Phys. A~ \textbf{34} (2001), 5969--5985.

\bibitem{ES}
P.~Exner and P.~{\v S}eba, \emph{Bound states in curved quantum waveguides},
  J.~Math.~Phys. \textbf{30} (1989), 2574--2580.

\bibitem{Fermi}
E.~Fermi, \emph{Sopra i fenomeni che avvengono in vicinanza di una linea
  oraria}, Atti R.~Accad. Lincei Rend. Cl. Sci. Fis. Mat. Natur. \textbf{31}
  (1922), 21--23, 51--52, 101--103. Also \emph{The Collected Works of
  E.~Fermi~1}, The University of Chicago Press, Chicago (1962), 17--19.

\bibitem{Fiala}
F.~Fiala, \emph{Le probl\`eme des isop\'erim\`etres sur les surfaces ouvertes
  \`a courbure positive}, Comment. Math. Helv. \textbf{13} (1940--41),
  293--346.

\bibitem{FrHe}
R.~Froese and I.~Herbst, \emph{Realizing holonomic constraints in classical and
  quantum mechanics}, Commun.~Math.~Phys. \textbf{220} (2001), 489--535.

\bibitem{GJ}
J.~Goldstone and R.~L. Jaffe, \emph{Bound states in twisting tubes}, Phys. Rev.
  B~ \textbf{45} (1992), 14100--14107.

\bibitem{Gray}
A.~Gray, \emph{Tubes}, Addison-Wesley Publishing Company, New York, 1990.

\bibitem{Hartman50}
P.~Hartman, \emph{On the local uniqueness of geodesics}, Amer.~J. Math.
  \textbf{72} (1950), 723--730.

\bibitem{Hartman64}
\bysame, \emph{Geodesic parallel coordinates in the large}, Amer.~J. Math.
  \textbf{86} (1964), 705--727.

\bibitem{Hartman-Wintner}
P.~Hartman and A.~Wintner, \emph{On the third fundamental form of a surface},
  Amer.~J. Math. \textbf{75} (1953), 298--334.

\bibitem{JK}
H.~Jensen and H.~Koppe, \emph{Quantum mechanics with constraints}, Ann.~Phys.
  \textbf{63} (1971), 586--591.

\bibitem{Kli}
W.~Klingenberg, \emph{A course in differential geometry}, Springer-Verlag, New
  York, 1978.

\bibitem{KJ}
H.~Koppe and H.~Jensen, \emph{Das prinzip von d'{A}lembert in der klassischen
  mechanik und in der quantentheorie}, Sitzungber. der Heidelberger Akad. der
  Wiss., Math.-Naturwiss. Klasse \textbf{5} (1971), 127--140.

\bibitem{LCM}
J.~T. Londergan, J.~P. Carini, and D.~P. Murdock, \emph{Binding and scattering
  in two-dimensional systems}, LNP, vol. m60, Springer, Berlin, 1999.

\bibitem{Mitchell}
K.~A. Mitchell, \emph{Gauge fields and extrapotentials in constrained quantum
  systems}, Phys. Rev.~A \textbf{63} (2001), art.~042112.

\bibitem{RS4}
M.~Reed and B.~Simon, \emph{Methods of modern mathematical physics}, vol. IV
  \emph{{A}nalysis of Operators}, Academic Press, New York, 1978.

\bibitem{RB}
W.~Renger and W.~Bulla, \emph{Existence of bound states in quantum waveguides
  under weak conditions}, Lett.~Math.~Phys. \textbf{35} (1995), 1--12.

\bibitem{Tol}
J.~Tolar, \emph{On a quantum mechanical d'{A}lembert principle}, Group
  theoretical methods in physics, LNP, vol. 313, Springer, 1988, pp.~268--274.

\end{thebibliography}
%
\providecommand{\bysame}{\leavevmode\hbox to3em{\hrulefill}\thinspace}
\providecommand{\MR}{\relax\ifhmode\unskip\space\fi MR }
\providecommand{\MRhref}[2]{%
  \href{http://www.ams.org/mathscinet-getitem?mr=#1}{#2}
}
\providecommand{\href}[2]{#2}

\end{document}